\documentclass[aps,pra,twocolumn,groupedaddress,amsmath,amssymb]{revtex4-2}
\usepackage{booktabs}
\usepackage{colortbl}
\usepackage{graphicx}% Include figure files
\usepackage{dcolumn}% Align table columns on decimal point
\usepackage{bm}% bold math
\usepackage{longtable}% long table
\usepackage{float}
\usepackage{siunitx}
\usepackage{textcomp}

\begin{document}

%\preprint{}

\title{Authentication of optical physical unclonable functions based on single-pixel detection}


\author{Pidong Wang}
 \email[]{wangpidong\_mtrc@caep.cn}
%\thanks{}
\author{Feiliang Chen}
\author{Dong Li}
\author{Song Sun}
\author{Feng Huang}
\author{Taiping Zhang}
\author{Qian Li}
\author{Kun Chen}
\author{Yongbiao Wan}
\author{Xiao Leng}
\author{Yao Yao}
 \email[]{yaoyao\_mtrc@caep.cn}


\affiliation{Microsystem and Terahertz Research Center, China Academy of Engineering Physics, Chengdu 610200, China}
\affiliation{Institute of Electronic Engineering, China Academy of Engineering Physics, Mianyang 621999, China}

\date{\today}

\begin{abstract}
Physical unclonable function (PUF) has been proposed as a promising and trustworthy solution to a variety of cryptographic applications. Here we propose a non-imaging based authentication scheme for optical PUFs materialized by random scattering media, in which the characteristic fingerprints of optical PUFs are extracted from stochastical fluctuations of the scattered light intensity with respect to laser challenges which are detected by a single-pixel detector. The randomness, uniqueness, unpredictability, and robustness of the extracted fingerprints are validated to be qualified for real authentication applications. By increasing the key length and improving the signal to noise ratio, the false accept rate of a fake PUF can be dramatically lowered to the order of $10^{-28}$. In comparison to the conventional laser-speckle-imaging based authentication with unique identity information obtained from textures of laser speckle patterns, this non-imaging scheme can be implemented at small speckle size bellowing the Nyquist--Shannon sampling criterion of the commonly used CCD or CMOS cameras, offering benefits in system miniaturization and immunity against reverse engineering attacks simultaneously.
\end{abstract}

%\keywords{}

\maketitle


\section{Introduction}
Physical unclonable functions (PUFs) are physical entities with unique security behaviors, which make them ideal in the field of hardware security \cite{gao2020physical,gianfelici2020theoretical,sakhare2020review,herder2014physical,4261134,RN1049}.
The uncontrollable nature of complex random disorders inside PUFs, which appear during the manufacturing process, makes duplication of them impractical or at such an expensive cost which renders it worthless \cite{shariati2010random,arapinis2019quantum}.
Optical PUF, composed of bulk random scattering media \cite{RN916,RN917,RN906,RN22,nikolopoulos2017continuous}, thin random scattering layer (such as regular paper \cite{nature1}) or rough scattering surface (such as laser-engraved surface \cite{shariati2010random,zhang2016robust}), is a subset of PUFs of particular interest.
The huge number, small size and random distribution of dense micro/nanostructures inside or on the surface of an optical PUF make it highly informative, resource-efficient, intrinsically tamper-resistant, inexpensive to fabricate, but difficult or too costly to duplicate \cite{RN916,RN917,Tuyls2007,RN906,Tuyls2007,Anderson:17,10.1117}. 
Moreover, with mature image-based feature extraction methods, the fingerprint information of optical PUFs could be easily accessed from textures of laser speckle patterns, which are derived from mutual interference of randomly scattered laser light from the surface and inside of optical PUFs and thus are especially favorable in 3D random physical feature exploration \cite{RN916,RN917,Shariati2012,Shariati2012a,Yeh:12,Liao2012,Costa2018}.
These perfect characters render optical PUF excellent in entity authentication.

Unfortunately, the development of reverse engineering techniques poses severe threat to optical PUFs. For example, nanoscale synchrotron X-ray computed tomography (SXCT) and 3D nanoprinting are now capable of achieving spatial resolution down to tens of nanometers and are still in rapid development \cite{RN1026,RN989,barner20173d,winkler2018high,winkler20193d,winkler2017direct}, which makes it in principle possible to accurately capture the precise nanostructure of an optical PUF and even reproduce it, especially for those composed of random rough surface. Thus it is necessary to increase the degree of complexity or cost of resources against the reverse engineering attacks on optical PUFs. 

Taking into consideration of the speed and resource cost of reverse engineering techniques, it is favorable to achieve the goal by simply increasing the effective volume of an optical PUF, namely increasing its actually illuminated region.
Since the thickness would affect both transmittance of laser light and angular sensitivity of laser speckle for bulk and layered scattering media \cite{RN948,Anderson_2015,RN1015,RN962,RN1012}, it is better to increase the effective volume via increasing the optical illumination area for optical PUFs composed of bulk and layered materials, just the same as optical PUFs composed of rough scattering surface. However, since the mean speckle size of a laser speckle pattern is proportional to $z/d$, where $z$ is the distance of the camera from the optical PUF and $d$ is the size of the laser illumination area \cite{RN982}, the speckle will be comparable to or even smaller than the pixel of commonly used CCD or CMOS cameras when the laser illumination area is large enough, resulting that the speckle pattern will be averaged out and can not be clearly observed \cite{stallinga2008method}. On the other hand, as system miniaturization is always on demand, the camera for laser speckle pattern observation is expected to be mounted as close to the optical PUF as possible, namely the value of $z$ is expected to be as small as possible, which would also lead to fuzzy imaging of laser speckle patterns. 
%Consequently, the implementation of an optical PUF is difficult because optical PUFs are image-based PUFs which rely on speckle pattern observation and processing \cite{RN916,RN917,RN948}. Therefore, non-image-based implementation of optical PUFs especially at small speckle size is imperative for highly secure integrated applications.
Consequently, it will be difficult to evaluate characteristic random features and extract identity information of an optical PUF from textures of laser speckle patterns, which would make conventional laser-speckle-imaging based authentication of optical PUFs impractical.
Therefore, a non-imaging based scheme applicable at small speckle size is imperative for highly secure integrated authentication of optical PUFs.

In this work, we propose a non-imaging based authentication scheme for optical PUFs executable at small speckle size, in which response signals of optical PUFs are detected by a single-pixel detector. We first theoretically and experimentally validate the feasibility of fingerprint information accession from single-pixel detected stochastical fluctuations of the scattered light with respect to challenges under the circumstance of small-sized laser speckles which are fuzzy to the commonly used CCD or CMOS cameras.
Randomness, uniqueness, unpredictability and robustness of the binary keys generated from the fluctuations are statistically evaluated for the necessary requirements of authentication applications.
The method for reduction of the false acceptance rate (FAR) and false rejection rate (FRR) in authentication is also discussed, as well as its positioning accuracy requirement.
At last an associated authentication protocol implementable at small speckle size is designed. 
Compared to the conventional laser-speckle-imaging based authentication with identity information extracted from textures of laser speckle patterns, such a non-imaging scheme leverages the advantage of small speckles and thus is advantageous in increasing both the security against reverse engineering attacks and the integration degree of the application units at the same time. 

\section{Authentication setup and principles}
\begin{figure}[b]
	\includegraphics[width=8.5cm]{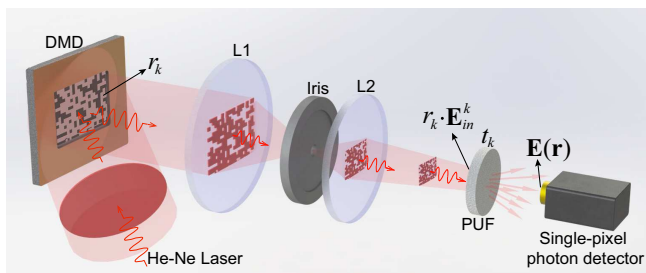}
	\caption{Experimental setup: a DMD creates challenges by randomly shaping the wavefront pattern of a He-Ne laser beam. After passing through the lens L1, iris and another lens L2, challenges are projected on the surface of the PUF in sequence. The scattered light in the far field of each challenge is detected by a single-pixel photon detector.}
	\label{fig1}
\end{figure}
A schematic for the authentication setup is shown in Fig. \ref{fig1}. An expanded and collimated Helium-Neon laser beam illuminates on a binary amplitude digital micro-mirror device (DMD) to encode the beam by shaping the laser wavefront with randomly configuring the reflection pattern of the DMD, namely randomly setting the reflection coefficient $r_k$ ($r_k=$ 0 or 1) of each DMD pixel.
In the methodology of PUF-based authentication protocols, such encoded wavefront corresponding to each DMD configuration is usually referred to as a ``challenge" \cite{RN916,RN22}.
A lens, L1, Fourier-transforms the encoded beam and an iris in its Fourier plane blocks all the diffraction orders except the -1st, which carries all the encoded information of the beam \cite{RN1055,Descloux:16,Gregory:88}. 
Passing through another lens L2, the encoded beam (i.e. challenge) is projected onto the surface of the PUF to stimulate it.
After complex interactions with the optical PUF, the scattered light is detected by a single-pixel photon detector in the farfield. 
We will show that the detected light, even at small speckle size, will fluctuate with the variation of challenges and can be treated as the response signal to the challenge in authentication applications.

The electric field of the scattered light at a point $\mathbf r$ on the surface of the detector can be calculated by \cite{RN981,RN984}
\begin{equation}\label{eq1}
\mathbf{E(r)}=\sum_{k=1}^{m}{t_k}\cdot r_k \cdot \mathbf{E}_{in}^k,
\end{equation}
where $k$ and $m$ are the index and the total number of DMD pixels respectively; $\mathbf E_{in}^k$ is the light electric field at the incident surface of the PUF; $r_k$ is the reflection coefficient of the k-th DMD pixel which follows an equiprobable Bernoulli distribution to get the maximum randomness (i.e. $r_k$ is set to be 0 or 1 with equal probability); $t_{\mathit k}\mathit=\mathit\vert t_{\mathit k}\mathit\vert\mathit\cdot e^{\mathit i{\mathit\phi}_{\mathit k}}$ describes the transfer of $\mathbf E_{in}^k$ to the point $\mathbf{r}$, where the amplitude $|t_k|$ is Rayleigh distributed and the phase $\phi_k$ is uniformly distributed in the interval $[-\pi,\pi]$ \cite{RN980,RN1039}. 

As the random nature of $t_k$, the electric field of the scattered light, as well as the corresponding light intensity, will fluctuate stochastically with random modulations of $r_k$.
Moreover, it could be noted that Eq. (\ref{eq1}) shares almost the same form as the function used by Goodman to calculate the electric field inside a single speckle pattern \cite{RN982}, except that the electric field of Eq. (\ref{eq1}) changes with the modulation of $r_k$ while the electric field inside a single speckle pattern varies with the observation point. Therefore, following the same derivation steps with Goodman \cite{RN982}, it can be concluded from Eq. (\ref{eq1}) that the fluctuation of the single-pixel detected intensity of the scattered light follows the gamma distribution:
\begin{equation}\label{eq2}
p_{\scriptscriptstyle I_\alpha}(I_\alpha)=\frac1{\Gamma(\mu)}\left(\frac\mu{\left\langle I_\alpha\right\rangle}\right)^\mu I_\alpha^{\mu-1}\exp\left(-\frac{\mu I_\alpha}{\left\langle I_\alpha\right\rangle}\right).
\end{equation}
Here $I_\alpha=\iint_{\mathbf{r}\in\alpha}|\mathbf{E}(\mathbf{r})|^2d\mathbf{r}$ is the integrated light intensity over the active area $\alpha$ of the detector. The angular brackets $\left<\cdots\right>$ denote an average over the ensemble of random challenges and thus $\left\langle I_\alpha\right\rangle$ is the mean intensity detected by the single-pixel detector.  
% $\left\langle I_\alpha\right\rangle=\left<\sum_{k=1}^{m} r_k\cdot \left|t_k\right|^2 \cdot \left|{\mathbf E}_{\mathit in}^k\right|^2\right>$.
$\Gamma(\mu)$ represents the gamma function.
$\mu$ is the shape parameter of the gamma distribution and its value depends on the size of the speckle: for speckles with average diameter larger than the size of the detector, $\mu\approx 1$, meaning that $I_\alpha$ approaches the negative exponential distribution, while in the opposite case, namely for speckles with average diameter smaller than the size of the detector, $\mu>1$, implying that $I_\alpha$ follows the gamma distribution.
As a result, the signal detected by the single-pixel detector will fluctuate stochastically with respect to the randomly encoded challenges no matter what the size of the speckle is, and the speckle size only affects the shape of the distribution.
%It can be noted that the distribution is the same as the intensity distribution inside a single speckle pattern, as well as the dependence of the value of $\mu$ on the size of the detector \cite{RN982}.

In our experiment we have access to the photon number of the scattered light. The number of photons $N$ detected by the single-pixel photon detector in a time interval of $\Delta t$ equals to $I_\alpha\cdot\Delta t/\hbar\omega$, where $\omega$ is the angular frequency of the challenge beam and $\hbar$ is the reduced Planck constant. Thus $N$ is also gamma-distributed, and its distribution can be written as
\begin{equation}\label{eq3}
\begin{aligned}
p_{\scriptscriptstyle N}(N)&=p_{\scriptscriptstyle I_\alpha}\left(N\hbar\omega/\Delta t\right)\cdot\frac\partial{\partial N}I_\alpha \\
&=\frac1{\Gamma(\mu)}\left(\frac\mu{\left\langle N\right\rangle}\right)^\mu N^{\mu-1}\exp\left(-\frac{\mu N}{\left\langle N\right\rangle}\right),
\end{aligned}
\end{equation}
where $\left<N\right>=\left<I_\alpha\right>\cdot\Delta t/\hbar \omega$ is the mean photon number. Thus the photon number $N$ follows the same fluctuation behavior as $I_{\alpha}$.

Equations \eqref{eq1}-\eqref{eq3} indicate that the detected signals ($I_\alpha$ and $N$) are related to both the physical features of the optical PUF and the parameters of the challenge light ($\left|\mathbf E_{in}^k\right|$ and $r_k$). To eliminate the impact of the challenge light intensity, $\left|\mathbf E_{in}^k\right|$ should keep approximately equal at each pixel and constant during the detection, besides, the number of reflecting DMD pixels ($\sum_{k=1}^{m}r_k $) should also be fixed. As a result, the fluctuation of the detected signal is only related to the binary reflection patterns displayed on the DMD, namely the encoded information, and the physical features of the PUF. 
Hence the fluctuation can be regarded as unique response (i.e. fingerprint) of a specific PUF to a specific challenge sequence, and based on it the authentication can be implemented.

Moreover, being the same with laser speckle, the fluctuation is also induced by mutual interference of randomly scattered light coming from micro/nanostructures both inside and on the surface of the optical PUF, thus it retains the advantage of laser speckle in easy exploration of 3D physical information.

\section{Experimental demonstration}
\subsection{Stochastical fluctuations of the scattered light and binary key generation}
Our optical PUFs are fabricated by titania (TiO$_2$) nanoparticles with an average grain size of \SI{200}{nm} randomly immersed in polymethy methacrylate (PMMA) films of thickness about \SI{17}{\mu m}. Their transport mean free path is measured to be approximately \SI{1.3}{\mu m} by coherent backscattering experiment \cite{wolf:jpa-00210675}, which implies that a photon will be scattered 13 times on average to transport through the film, leading the PUF to be a multiple scattering media \cite{ishimaru1978wave,meretska2017analytical,ott2010quantum}. The diameter of the incident beam is about 4 cm, which is large enough to cover the full active area of the DMD (Vialux V-7001). The beam reflected by the DMD encoding area is rectangular in shape with a length-to-width ratio of 4:3. 
Figure 2(a) plots the mean speckle size $D$ versus the distance $z$ between the optical PUF and the CCD camera (Thorlabs 340M-USB, with a pixel size of $\SI{7.4}{\mu m}\times\SI{7.4}{\mu m}$), as well as the width of the incident beam $d$, which is controlled by adjusting the width of the encoding area of the DMD (i.e. the width of the DMD reflection pattern).
The size around the white dashed line is about twice the camera's pixel size, which is the minimum Nyquist--Shannon sampling size. In the area below the white dashed line, the speckle size is evaluated by calculating the width of the normalized autocorrelation function of the intensity pattern of the speckle \cite{RN1032,RN1034}, while in the area above, it is evaluated by surface extrapolating from data below the white dashed line using the formula $D\propto z/d $ \cite{RN982}, owing to the fact that the speckles in this area are so small that they could not be observed clearly by the camera. It can be seen that the mean size of the speckle increase with $z$ and decrease with $d$.
%Fig. 2(b) shows a speckle pattern recorded at point O ($z=35\ mm$ and $d=0.66\ mm$), in which speckles can be observed clearly by the camera and are of moderate size for speckle-pattern-based authentication. Speckles in Fig. 2(c) are of almost the same mean size with that in Fig. 2(b), which indicates that the distance of an optical PUF from the camera should be increased simultaneously with the increase of the illumination area.
\begin{figure}[!t]
	\centering\includegraphics[width=86mm]{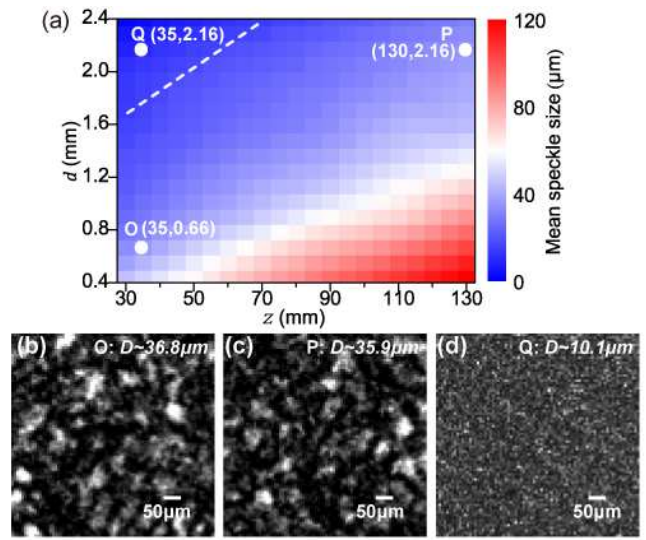}
	\caption{(a) Mean speckle size $D$ versus $z$ and $d$, which are the distance from the optical PUF to the CCD camera and the width of the rectangular incident challenge beam respectively. White dashed line corresponds to the minimum Nyquist--Shannon sampling size. (b)(c)(d) Speckle patterns observed at points O, P and Q respectively.}
	\label{fig2}
\end{figure}

For illustration purpose, Figs. 2(b) and 2(c) show two speckle patterns recorded at points O ($z=\SI{35}{mm}$ and $d=\SI{0.66}{mm}$) and P ($z=\SI{130}{mm}$ and $d=\SI{2.16}{mm}$) respectively, inside which textures of laser speckles can be observed clearly and are of moderate speckle size for laser-speckle-imaging based authentication.
The almost same mean speckle size in these two figures indicates that the camera's distance to the optical PUF should be increased with the illumination area simultaneously in order that the mean speckle size remains unchanged, implying that we cannot increase the effective volume of an optical PUF without changing the size of a laser-speckle-imaging based authentication system.
Figure 2(d) records a speckle pattern at point Q ($z=\SI{35}{mm}$ and $d=\SI{2.16}{mm}$), which is of the same camera's distance to the optical PUF as point O and of the same illumination area as point P, and inside the figure speckles are of small mean size of only approximately 1.36 times the camera pixel size, which is much smaller than the minimum Nyquist--Shannon sampling size.
Point Q has both large illumination area and short camera's distance to the optical PUF, nevertheless, textures of the speckle patterns at this point are so fuzzy to the camera that they are not applicable for laser-speckle-imaging based authentication.
%Comparing with points O and P, point Q is of large illumination area and of short distance between the camera and the PUF simultaneously.
\begin{figure}[t]
	\centering\includegraphics[width=86mm]{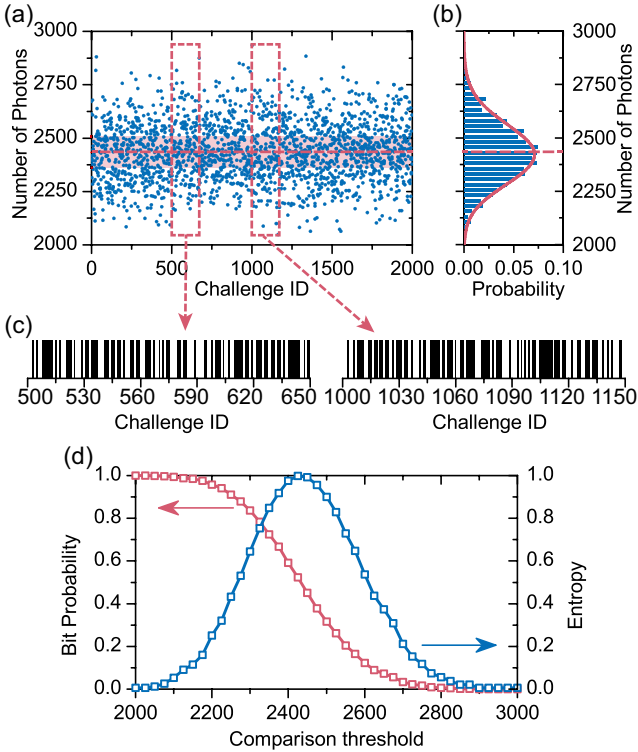}
	\caption{(a) Number of photons detected by the single-pixel detector when illuminating a PUF with randomly modulated challenges during a time interval of \SI{1}{ms}. (b) Distribution of the detected photons. (c) Two 150-bit binary barcode key tags. (d) The probability of each of the bits being set to 1 (red line), calculated by averaging over an ensemble of 1000 bits of different encoded challenges. The blue line graphs the Shannon entropy.}
	\label{fig3}
\end{figure}

To detect the fuzzy speckle response at point Q, the camera is replaced by a single-pixel photon detector with an active area of \SI{180}{\mu m} in diameter (Excelitas SPCM-AQRH-14). When a photon is detected, the detector will output a TTL pulse, and we exploit a data acquisition card (NI USB-6351) to count the number of the outputted TTL pulses, i.e., the number of photons.
The DMD encoding area consisting of $960\times720$ pixels is divided into $40\times30$ segments to shape the beam wavefront, each of which consists of $24\times24$ pixels and acts as a superpixel for amplitude modulation of the incident wavefront.
In order to get a challenge of a fixed amount of reflecting segments while with the maximum randomness, half of the segments are randomly chosen and set to 1, allowing a maximum of $\dbinom{1200}{600}\approx 10^{360}$ distinguishable challenges. The Shannon Entropy of each bit in a challenge equals to $-\frac{1}{2}\log_2 (\frac{1}{2})-\frac{1}{2}\log_2 (\frac{1}{2})=\SI{1}{bit}$, which is the maximum Entropy of a binary bit, indicating that the challenges are incompressible and hard to be predicted \cite{entropy1,entropy2}. Fluctuations of the scattered photon numbers are subsequently measured by the single-pixel photon detector when illuminating a PUF with randomly encoded challenges, as shown in Fig. 3(a). Figure 3(b) plots the statistical distribution of the detected photon numbers, and the red line is the curve fitting with Eq. (\ref{eq3}), with $\left<N\right>=2429$ and $\mu=301.8$.
Based on the analysis in Section II, the value of $\mu\gg1$ indicates that the speckle is much smaller than the detector, just according with the real size of the speckle (\SI{10.1}{\mu m}) and the detector (\SI{180}{\mu m}).

By comparing with a comparison threshold, binary keys can be generated from the detected fluctuating signals.
Figure 3(c) shows two 1D 150-bit barcode key tags generated from points in the red dashed box in Fig. 3(a) with the comparison threshold equaling to the median value of the detected photon numbers, i.e. 2435, as the red dashed line plots in Figs. 3(a) and 3(b).
Figure 3(d) presents the influence of the comparison threshold value, in which the average probability of a bit being set to 1 equals to 0.5 and the Shannon entropy \cite{RN917} reaches its maximum when the comparison threshold approaches the median value of the detected fluctuating photon numbers, implying that binary numbers with bitwise maximum entropy are produced at this point. 
Therefore, in the ensuing discussion, the comparison threshold for binary key generation is set to the median value $N_m$ of the detected response signals of each PUF.

\subsection{Statistical properties of the generated binary keys}
To enable the usage of optical PUFs with single-pixel detection of response signals in authentication applications, it is necessary to carefully evaluate randomness, uniqueness, unpredictability, and robustness of the detected response signals \cite{gianfelici2020theoretical,herder2014physical}.

%\subsubsection{Randomness evaluation}
\begin{table*}[ht]
	%\small
	%	\footnotesize
	\caption{Results of the NIST randomness test}%\textbf{Results of the NIST randomness test}}
	%\vspace{0pt}
	%\centering
	\begin{ruledtabular}
		\begin{tabular}{lll}
			%\arrayrulecolor{tabcolor}
			%\toprule[1.5pt]
			%		\hline
			{Test type}                         & {Uniformity   of P-value}        &{Pass   rate (\%)}         \\
			%		\hline
			\midrule[0.4pt]
			Frequency                           & 0.554                            & 100                       \\
			Frequency   test within a block     & 0.616                            & 98                        \\
			Runs                                & 0.290                            & 97                        \\
			Longest   run of ones in a block    & 0.494                            & 97                        \\
			Binary   Matrix rank                & 0.213                            & 100                       \\
			Discrete   Fourier transform        & 0.616                            & 99                        \\
			Non   overlapping template matching & 0.0026\footnotemark[1]           & 96                        \\
			Overlapping   template matching     & 0.679                            & 99                        \\
			Universal                           & 0.740                            & 99                        \\
			Linear   complexity                 & 0.038                            & 99                        \\
			Serial                              & 0.419                            & 99                        \\
			Serial                              & 0.924                            & 99                        \\
			Approximate   entropy               & 0.052                            & 99                        \\
			Cumulative   sums –forward          & 0.350                            & 100                       \\
			Cumulative   sums –backward         & 0.437                            & 100                       \\
			Random excursions                   & 0.052\footnotemark[1]            & 97.67                     \\
			Random   excursions variant         & 0.027\footnotemark[1]            & 97.67                     \\
			%		\hline
			%\bottomrule[1.5pt]
		\end{tabular}\\
	\end{ruledtabular}
	\footnotetext[1]{More than one value is obtained and the values given for these tests are their minimum.}
\end{table*}

Here the randomness of binary numbers generated from the stochastically fluctuating photon numbers detected at point Q in Fig. 2(a) is tested by a standard national institute of standards and technology (NIST) randomness test suite \cite{RN1036}. The length of the used binary number sequence for the first 15 benchmarks of NIST test (i.e. all except “Random excursions” and “Random excursions variant”) is 10000, while the length for the last 2 benchmarks (i.e. “Random excursions” and “Random excursions variant”) is 1000000. Each of the NIST test benchmarks is repeated 100 times. Table 1 shows the results of the NIST test, and it confirms that the generated binary numbers successfully pass the NIST randomness tests, i.e., the probability value (P-value) is greater than 0.01 and the uniformity exceeds 0.0001 \cite{RN1036,RN1025,RN48}. We also evaluate the randomness of binary numbers generated from Fig. 2(d) by comparing the intensity at each pixel with the median intensity of the whole picture, but they fail to pass the NIST test, because the intensities between the adjacent pixels are correlated \cite{RN917,feng1988correlations}. Consequently, if parallel detection with multiple detectors is used to improve the generation speed and efficiency of binary numbers, as well as to reduce the number of challenges required for the generation of binary keys of a fixed length, detectors should be mounted far enough between each other.

\begin{figure}[b]
	\centering\includegraphics[width=86mm]{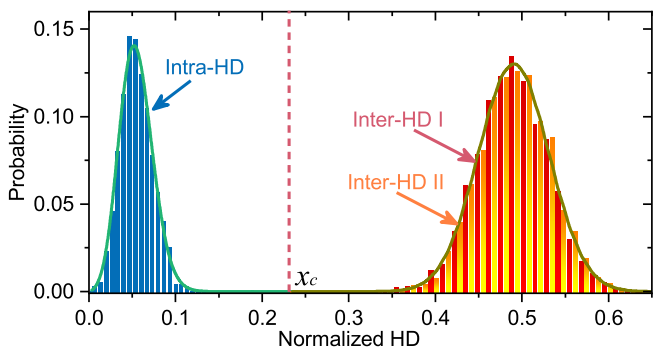}
	\caption{Distribution of the normalized HD among binary keys: Intra-HD (in blue), Inter-HD I (in red), Inter-HD II (in orange). Green line: fitting curve of Intra-HD. Olive line: fitting curve of two Inter HDs. $x_c$: cross point of the two fitting curves.}
	\label{fig4}
\end{figure}

%\subsubsection{Uniqueness, unpredictability and robustness evaluation}
To evaluate the uniqueness, unpredictability and robustness of the detected response signal, binary keys of a fixed length of $L$ bits generated from the stochastically fluctuating photon numbers detected at point Q in Fig. 2(a) are statistically analyzed by calculating the normalized Hamming distance (HD) among them, which counts the percentage of bits that differ between two keys. 
Here two kinds of inter HDs are calculated: inter HDs among keys generated by illuminating different PUFs with the same challenge sequence (Inter-HD I), and inter HDs among keys generated by illuminating the same PUF with different challenge sequences (Inter-HD II). Inter-HD I assesses the uniqueness of a binary key to its generating PUF, while Inter-HD II evaluates the uniqueness of a binary key to its generating challenge sequence and the unpredictability of the binary key. The red and orange histograms in Fig. 4 plot the distributions of Inter-HD I and Inter-HD II respectively, both of which are calculated among 50 150-bit binary keys. The two inter HDs share the same distribution and both can be well fitted by a modified binomial distribution:
\begin{equation}\label{eq4}
f(x)=\binom{L}{\left[ L\cdot x\right]}p^{\left[ L\cdot x\right]}(1-p)^{L-{\left[ L\cdot x\right]}},
\end{equation}
as the olive fitting curve in Fig. 4 shows. Here $x$ is the normalized HD and the rounded number $\left[ L\cdot x\right]$ represents the number of bits that differ in two different keys. The parameter $p$ represents the mean normalized HD. 
The resulting distributions of the two inter HDs have the same mean of approximate 0.496, which means an almost even odds of a bit to be different, implying that a binary key is unique not only to its generating PUF but also to the challenge sequence and it is unpredictable. The variances of the two histograms equal to $1.67\times 10^{-3}$, so there are $0.496\times(1-0.496)/1.67\times 10^{-3}\approx150$ independent identically distributed variables in the binary key, which represents that all the bits in the binary key are independent identically distributed and a theoretical key space size on the order of $2^L$ distinguishable $L$-bit binary keys can be provided by the PUF.

The robustness of the generated binary keys is estimated with HDs among remeasured copies of a binary key, labeled as Intra-HD. The blue histograms in Fig. 4 show the distribution of the Intra-HD among 50 remeasured copies of a 150-bit binary key, and it can be well fitted by the modified binomial distribution [Eq. (\ref{eq4})], as indicated by the green fitting curve. The mean value of Intra-HD is 0.056 and its variance is $3.6\times10^{-4}$. The nonzero mean value demonstrates that the same PUF even mounted at the same place and illuminated by the same challenge sequence rarely produce completely identical binary keys. This is caused by system noises, including photon shot noise, intensity noise of the laser, dark current noise of the photon detector, light pollution in the environment, etc. Nevertheless, the almost zero mean distance indicates that there are only a few bits (about 8 bits on average) differ in their remeasured copies, thus the binary key is robust and retrievable. 

As the random, unique, unpredictable and robust nature, the generated binary keys can be regarded as fingerprints to optical PUFs and are qualified for authentication applications. Moreover, due to the large space size of challenge and response and the uniqueness of a binary key to its corresponding challenge sequence, it is possible to use each binary key only once, which could greatly improve the security of the authentication \cite{RN916}.

\subsection{Authentication with the generated binary keys}
Based upon the above statistical analysis, authentication can be implemented. To decide whether a candidate PUF is the right one previously enrolled in the database in authentication, the minimum probability-of-error decision rule is to reject the candidate when the probability that the two PUFs are the same is less than the probability that they are different \cite{RN916}. The two fitting curves in Fig. 4 intersect at $x_c\approx0.221$, so we can arrive at a decision rule to reject a PUF’s authenticity if the $L$-bit binary key differ by more than $x_c \cdot L=0.221\times 150\approx 33$ bits. 

Due to the fact that point Q in Fig. 2(a) has both large illumination area and short detector's distance to the optical PUF, authentication based on single-pixel detection of response signals at this point is performed with a large effective volume of the optical PUF and a high integration level of the authentication system at the same time. And further it can be expected that the authentication can still be implemented even with the effective volume expanding to a far larger extent and the detector mounted far closer to the optical PUF than point Q, hence the single-pixel-detection based authentication can be implemented far more safely and compactly in respect to conventional laser-speckle-imaging based authentication.

\begin{figure*}[tp]
	\centering\includegraphics[width=13.5cm]{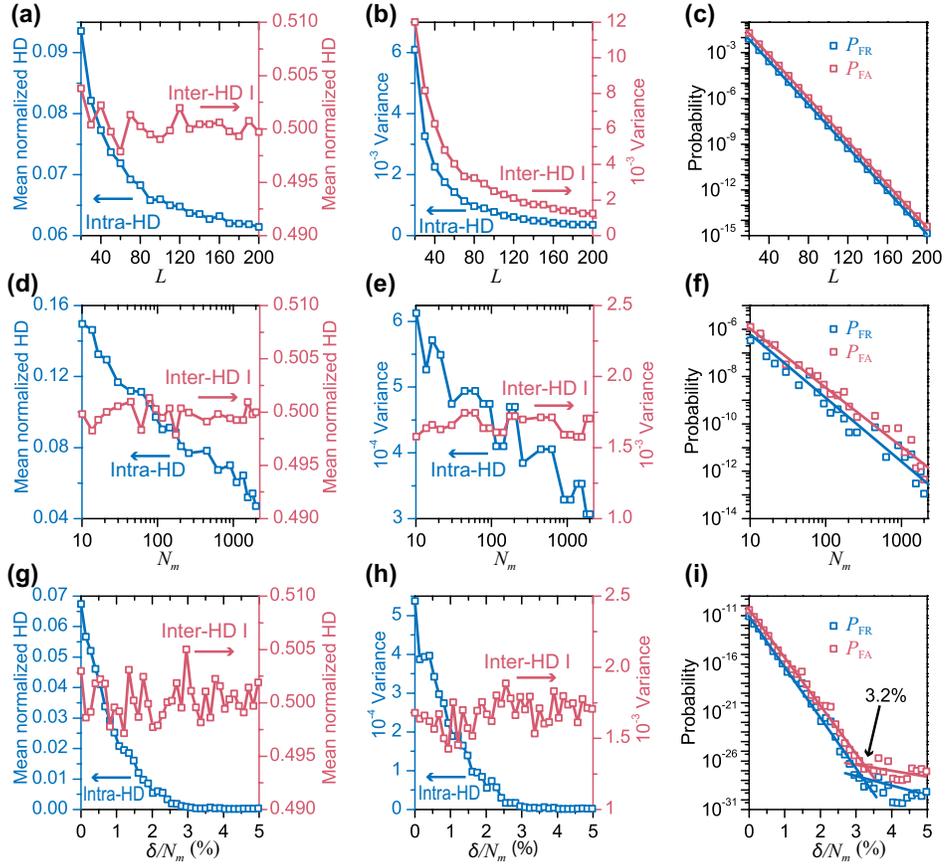}
	\caption{(a)(b)(c) Mean normalized HD, variance, FAR and FRR versus the key length respectively ($N_m=2429, \delta=0$). (d)(e)(f) Mean normalized HD, variance, FAR and FRR versus the $N_m$ respectively ($L=150, \delta=0$). (g)(h)(i) Mean normalized HD, variance, FAR and FRR versus the value of $\delta$ respectively ($L=150, N_m=2429$). Red and blue line in (c)(f)(i): fitting curves of FAR and FRR respectively.}
	\label{fig5}
\end{figure*}

\subsection{Reduction of FAR and FRR}
Since there is always a small overlap between the distribution of Inter-HDs and Intra-HDs around the intersection point $x_c$, false acceptance of a fake PUF and false rejection of a true PUF might occur, especially for cases of HDs close to $x_c$. %Therefore, FAR and FRR, which count the probabilities of the false acceptance and the false rejection respectively, are necessary for the estimation of the authentication result \cite{RN916}.
From this perspective, the authentication of a candidate PUF can be viewed as a fundamental problem of statistical inference, that is, binary hypothesis testing (i.e., true or fake of a candidate PUF) \cite{kay1998statistical}. In this context, FRR and FAR (also known as type I and II errors in statistical inference \cite{kay1998statistical}), which count the probabilities of the false rejection and the false acceptance respectively, are necessary to estimate the security of the authentication result \cite{RN916}.
FAR and FRR can be calculated by the following cumulative function:
\begin{equation}\label{eq5}
\begin{cases}
P_{\scriptscriptstyle F\!R}=1-F\left( \left[ L\cdot x_c \right] ,L,p_1\right) \\
P_{\scriptscriptstyle F\!A}=F\left( \left[ L\cdot x_c \right] ,L,p_2\right)
\end{cases}.
\end{equation}
Here $P_{\scriptscriptstyle F\!R}$ and $P_{\scriptscriptstyle F\!A}$ represent FRR and FAR respectively.  
$F\left(\left[L \cdot x_c \right],L,p_i\right)=\sum_{j=0}^{\left[L \cdot x_c \right]}\binom{L}{j}p_i^j(1-p_i)^{L-j},\ i=1\ \text{or}\ 2$,  which is the cumulative distribution function of the binomial distribution. The parameters $p_1$ and $p_2$ are the mean values of Intra-HD and Inter-HD I respectively. Therefore, we can arrive at a FAR of $6.3\times 10^{-12}$ and a FRR of $2.1\times 10^{-12}$ in the case of Fig. 4.

In real applications, it is important to reduce the FAR and FRR for the security of authentication. The statistical results in Fig. 4 illustrate that there are two effective ways to reduce FAR and FRR: reducing the variance of each distribution, and increasing the difference between $p_1$ and $p_2$.

As the variance of the normalized HD equals to $p_i (1-p_i)/L$  \cite{RN916}, $i=1\ \text{or}\ 2$, FAR and FRR can be lowered by increasing the key length $L$. Figures 5(a) and 5(b) plot the influence of the key length $L$ on the mean values of the normalized HDs and their corresponding variances, which illustrate that not only the mean Intra-HD $p_1$ but also the variances of both histograms are successfully decreased. Figure 5(c) shows that FAR and FRR are successfully lowered when increasing the key length, and they both can be well approximated by a simple decaying exponential.

As mentioned above, the mean Intra-HD $p_1$ is affected by many kinds of noises, hence it is possible to increase the signal to noise ratio (SNR) of the detected signal to lower the value of $p_1$ so as to increase the difference between $p_1$ and $p_2$. An effective and commonly used method to increase SNR is to increase the strength of the signal. To adopt this method here, the generated binary keys should be invariant to the intensity of the scattered light. The median value (i.e. comparison threshold for binary key generation) of the detected photon numbers $N_m$ can be calculated by the following equation:
\begin{equation}\label{eq6}
\begin{aligned}
0.5&=\sum_{N=0}^{N_m}p_{\scriptscriptstyle N}(N) \\
&=\frac{1}{\Gamma(\mu)}\mu ^\mu \sum_{\frac{N}{\left<N\right>}=0}^{\frac{N_m}{\left<N\right>}}\left(\frac{N}{\left<N\right>}\right)^{\mu-1}\cdot \exp \left(-\mu \cdot \frac{N}{\left<N\right>}\right),
\end{aligned}
\end{equation}
which implies that $N_m/\left<N\right>$ is a constant irrelevant to the value of $\left<N\right>$. 
The ratio of the detected signal to its mean value can be calculated by $N/\left<N\right>=\iint_{\alpha} \sum_{k=1}^{m} r_k\cdot \left|t_k\right|^2 \cdot d\mathbf{r}/ \left<  \iint_{\alpha} \sum_{k=1}^{m} r_k\cdot \left|t_k\right|^2 \cdot d\mathbf{r} \right>$, which is also a constant value for each challenge. So $N/N_m$ would be a constant value for each challenge under arbitrary mean scattering intensity, as well as under arbitrary median scattering intensity, as long as the mean and median intensity keep invariant during the key production. 
Therefore, increasing the intensity of the scattered light will not affect the binary key generation and can be used to increase the SNR here. Figures 5(d) and 5(e) plot the change of the mean normalized HDs and their corresponding variances with respect to the median value of the detected photon numbers. It can be seen that not only the value of $p_1$ but also its corresponding variance are successfully decreased. Figure 5(f) shows that FAR and FRR are successfully lowered with increasing $N_m$, and they can be well approximated by a decaying polynomial.

\begin{figure}[b]
	\centering\includegraphics[width=86mm]{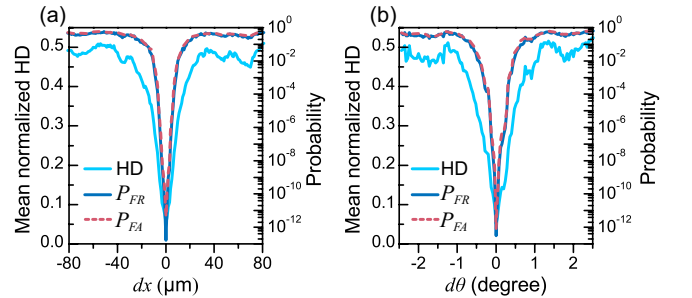}
	\caption{ (a) Translational sensitivity of the key. (b) Rotational sensitivity of the key. $L=150, N_m=2429, \delta=0$.}
	\label{fig6}
\end{figure}
\begin{figure*}[t]
	\centering\includegraphics[width=13.5cm]{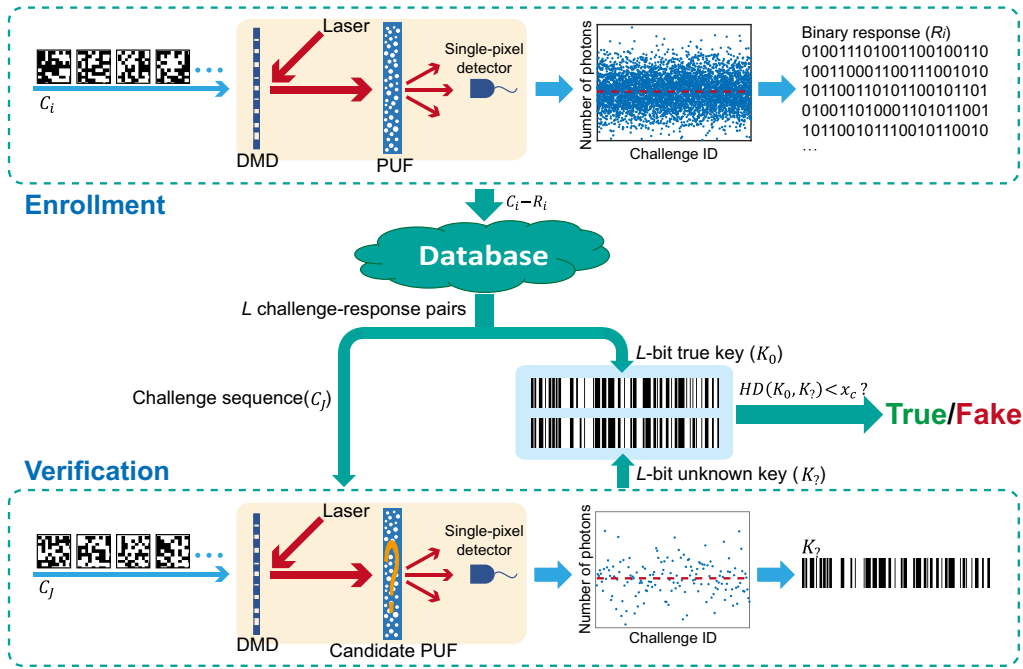}
	\caption{Authentication protocol based on single-pixel detection of response signals.}
	\label{fig7}
\end{figure*}

Another way to reduce the impact of noise is to discard those challenges with scattering intensities lie close to the comparison threshold, thus only response signals with high SNR are remained and stored. As indicated in Fig. 3(a), points inside the red area (i.e. points between $N_m-\delta$ and $N_m+\delta$, where $\delta$ is half the height of the red area) are discarded. Figures 5(g) and 5(h) plot the change of the mean normalized HDs and their corresponding variances with the value of $\delta$, in which the value of $p_1$ and its corresponding variance are also successfully decreased with increasing the value of $\delta$. The FAR and FRR also decrease approximately exponentially as shown in Fig. 5(i), but there is a threshold around $0.032N_m$ for $\delta$, where the decay rate at $\delta>0.032N_m$ is smaller than that at $\delta<0.032N_m$. This is caused by the elimination of all of the noise except the photon shot noise.
After optimization by increasing the SNR, including increasing the intensity of the scattered light and discarding responses with low SNR, the security parameter FAR can be lowered even to the order of $10^{-28}$ for authentication with a key length of 150 bits.

Based on the theoretical analysis in Secction II, the response signal detected by the single-pixel detector will fluctuate no matter what the size of the speckle is, and the speckle size only affect the line shape of the statistical distribution of the fluctuation, so the single-pixel detection scheme would still be feasible even for speckle pattern with speckle size of sub-micron or sub-wavelength. However, our present experiment shows that the fluctuation will be overwhelmed by noise with continuing reduction of the speckle size to such extent, so seeking methods to lower the SNR is necessary to validate the scheme in sub-micron and sub-wavelength region.

\subsection{Positioning accuracy requirement of optical PUF}
In authentication applications, optical PUFs would be repeatedly removed and reinserted into the light path, so the plug and play mount for the optical PUF should own a precise repeatability of position for the retrieval of response signals \cite{RN916,RN22,RN917,Anderson_2015}. In Figs. 6(a) and 6(b) the laser beam are successively translated and rotated about the PUF by a small amount of $dx$ and $d\theta$ respectively, which shows that a translation of approximate \SI{30}{\mu m} or a rotation of approximate \SI{0.5}{\degree} will cause the generated binary keys to decorrelate completely. Therefore, commercial products, such as Thorlabs KB1X1, whose position error is smaller than \SI{2.5}{\mu m} and \SI{26.72}{\mu rad}, is precise enough for the positioning of optical PUFs in authentication applications.

\section{Authentication protocol}
Given the preceding analysis, the generated binary keys, extracted from single-pixel detection of small-sized laser speckle response, are random, unique, unpredictable and robust enough for authentication applications. A simple authentication protocol is shown in Fig. 7, in which the execution can be divided into two steps: enrollment and verification:\\
\textbf{Enrollment stage.} A large number of challenges ($C_i$) randomly encoded by the DMD are sent to illuminate an optical PUF and their corresponding scattered light are detected by a single-pixel detector. Comparing with the median value, the detected signals are transformed into binary numbers ($R_i$). Challenges together with their corresponding binary responses are recorded to form challenge-response pairs ($C_i-R_i$), and then stored into a challenge-response database.\\
\textbf{Verification stage.} $L$ challenge-response pairs are randomly chosen from the challenge-response database to form a challenge sequence ($C_J$) and a $L$-bit true key ($K_0$). Similar to the enrollment stage, the candidate optical PUF is illuminated by this challenge sequence successively and the scattered light is detected by the single-pixel detector, after that the detected signals are transformed into a $L$-bit unknown key ($K_?$) by comparing them with their median value. Next the normalized HD between $K_0$ and $K_?$ is calculated. If the normalized HD is below a preset threshold $x_c$, the candidate optical PUF is authenticated. In order to obtain a safer execution of authentication, all used challenge-response pairs will be discarded based on the principle of one-time pad.

Moreover, as previously discussed, parallel detection with multiple detectors can be used in the authentication protocol to improve the binary key generation speed and efficiency, as well as to reduce the number of challenges required for each binary key, on the condition that multiple detectors are mounted far enough between each other.

\section{Conclusion}
In summary, we have presented a non-imaging based scheme for authentication of optical PUFs executable at small speckle size with fingerprint physical information detected by a single-pixel detector.
We theoretically and experimentally demonstrated that the intensity of the scattered light, with small-sized laser speckles fuzzy to the commonly used CCD or CMOS cameras, would fluctuate stochastically with the randomly encoded challenges. 
Binary keys obtained from the fluctuations are random, unique, unpredictable and robust enough to be qualified for authentication applications. 
Authentication with these binary keys has been displayed and three methods have been illustrated to reduce the FRR and FAR dramatically, including increasing the key length, increasing the intensity of the scattered light and discarding challenges with low SNR,   
which greatly enhanced the security of the authentication since the most important security parameter FAR can be lowered even to the order of $10^{-28}$. Moreover, it is worth noting that our scheme is a universal method which can be implemented with other types of single-pixel detector, in other spectral region, and even at smaller speckle size than that analyzed here (see Appendix \ref{A1} in detail).

Finally, a simple authentication protocol has also been designed, which is feasible at small speckle size.
In comparison with the conventional laser-speckle-imaging based authentication with identity information extracted from textures of laser speckle patterns, the designed authentication protocol not only retains the advantage of laser speckle in easy exploration of characteristic physical features both inside and on the surface of optical PUFs, but also offers advantages in increasing the complexity and costs on reverse engineering of the optical PUF while simultaneously increasing the integration level of the application units, which are beneficial for the safety and compactness in real applications.

\section*{Acknowledgments}
We gratefully acknowledge the anonymous referee for helpful comments and constructive suggestions. This work was supported by Science Challenge Project (TZ2018003-3) and National Natural Science Foundation of China (61875178, 12175204, 61805218,
12104423).
%%%%%%%%%%%%%%%%%%%%%%%%%%%%%%%%%%%%%%%%%%%%%%%%%%%%%%%%%%%
\appendix
\section{Universality of the single-pixel detection based authentication scheme}\label{A1}
In this appendix we demonstrate the universality of the proposed scheme, including implementations with other types of single-pixel detector, in other spectral region, and even beyond the smallest speckle size observable by the commercial cameras.
\begin{figure}[b]
	\centering\includegraphics[width=8.6cm]{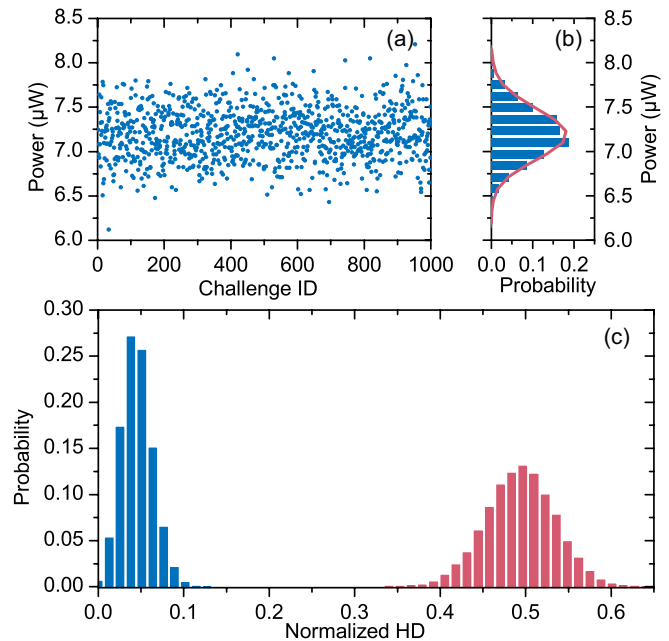}
	\caption{(a) Intensity fluctuation of the scattered light detected by the power meter when illuminating the PUF with randomly modulated challenges. (b) Distribution of the detected intensities (blue bars) and the gamma fitting curve (red line). (c) Distribution of Intra-HD (in blue) and Inter-HD I (in red). Laser wavelength: $\SI{632.8}{nm}$ (He-Ne laser).}
	\label{fig8}
\end{figure}

Other detectors are also feasible for the proposed single-pixel detection scheme. Figure~\ref{fig8} shows the experimental results with the single-pixel photon detector replaced by a Si photodiode power meter (Thorlabs S130C) located behind a pinhole ($\SI{200}{\mu m}$ in diameter). The results are consistent with what observed by the single-pixel photon detector.

\begin{figure}[t]
	\centering\includegraphics[width=8.6cm]{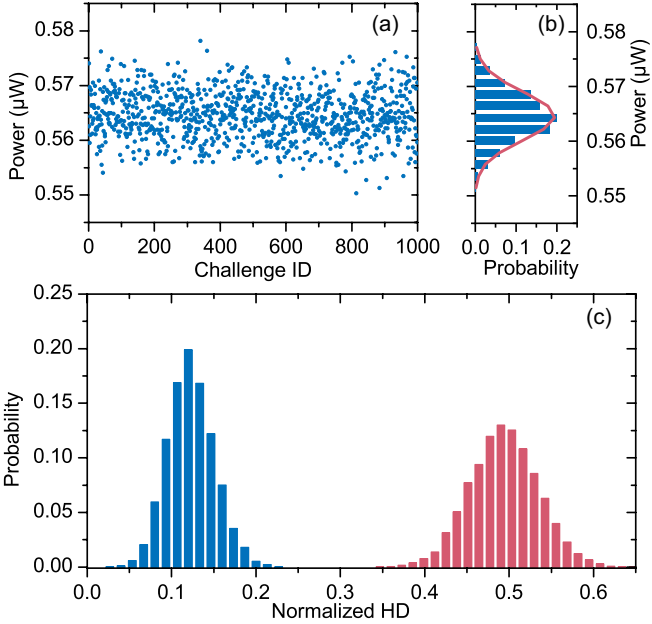}
	\caption{ (a) Intensity fluctuation of the scattered light detected by the power meter when illuminating the PUF with randomly modulated challenges. (b) Distribution of the detected intensities (blue bars) and the fitting curve with gamma function (red line). (c) Distribution of Intra-HD (in blue) and Inter-HD I (in red). Laser wavelength: $\SI{1550}{nm}$.}
	\label{fig9}
\end{figure}
The proposed single-pixel detection technique can also be utilized in other spectral region. Figure~\ref{fig9} shows the experimental results with He-Ne laser replaced by a telecom-band laser of $\SI{1550}{nm}$ and the single-pixel photon detector replaced by a Si photodiode based infrared power meter (Thorlabs S132C) located behind a pinhole of diameter $\SI{200}{\mu m}$, which are consistent with the results observed in visible region of wavelength $\SI{632.8}{nm}$.

\begin{figure}[h]
	\centering\includegraphics[width=8.3cm]{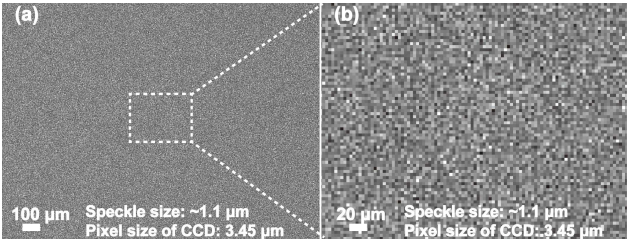}
	\caption{(a) Speckle pattern of speckle size about $\SI{1.1}{\mu m}$ recorded by Point Gray CCD camera. (b) A magnification of the image in the dashed box in (a).}
	\label{fig10}
\end{figure}
\begin{figure}[h]
	\centering\includegraphics[width=8.6cm]{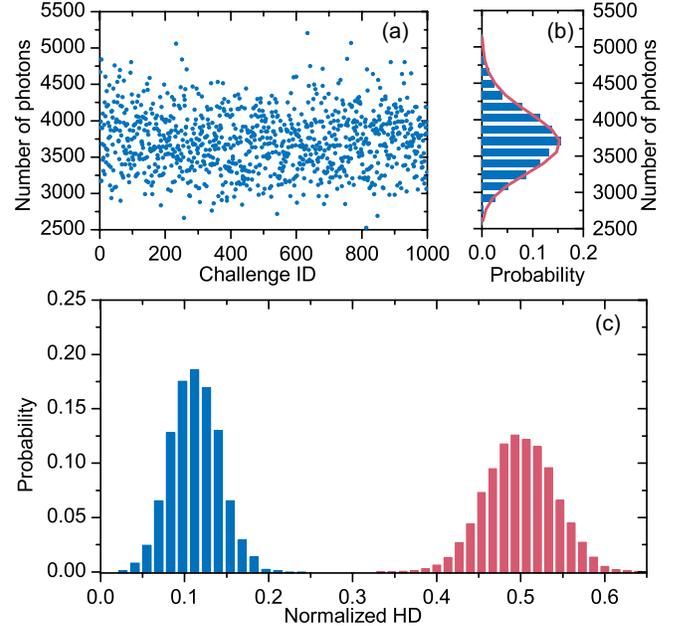}
	\caption{ (a) Number of photons detected by the single-pixel photon detector when illuminating the PUF with randomly modulated challenges. (b) Distribution of the detected photons (blue bars) and the fitting curve with gamma function (red line). (c) Distribution of Intra-HD (in blue) and Inter-HD I (in red).}
	\label{fig11}
\end{figure}

To our knowledge, commercial image sensor with the smallest pixel size is Samsung’s ISOCELL Slim GH1 whose pixel size is $\SI{0.7}{\mu m}$. Using this image sensor, the smallest speckle of size $\SI{1.4}{\mu m}$ can be observed according to the Nyquist-Shannon sampling theorem, while for speckles smaller than $\SI{1.4}{\mu m}$, fuzzy imaging will occur. We have performed experiment with speckle size smaller than $\SI{1.4}{\mu m}$, which shows that the proposed non-imaging scheme is still effective. Figure~\ref{fig10} presents the speckle pattern of size about $\SI{1.1}{\mu m}$, inside which the speckle spot can not be observed clearly. This image is recorded by a CCD camera of pixel size $\SI{3.45}{\mu m}$ (Point Gray GS3-U3-51S5M-C), which is the smallest pixel size we can access (even though the pixel size of the camera on many commonly used mobile phone is about $\SI{1}{\mu m}$, but the imaging lens integrated with the image sensor would affect the observation of such small speckles, so we don’t use the mobile phone camera). The speckle size of $\SI{1.1}{\mu m}$ is obtained by extrapolation with the method introduced in Section II. Figure~\ref{fig11}(a) plots the fluctuation of the fuzzy speckle response in Fig.~\ref{fig10} which is measured by a single-pixel photon detector (Thorlabs SPCM20A), and Fig.~\ref{fig11}(b) displays its statistical distribution. Figure~\ref{fig11}(c) shows the distribution of Intra-HD and Inter-HD I. Figure~\ref{fig11} demonstrates that this non-imaging authentication scheme is still effective for speckle patterns which is fuzzy to the commercial image sensor of the highest resolution (i.e. of the smallest pixel size) at present.

\end{document}